\documentclass[aps,prb,a4,reprint,twoside,floatfix,]{revtex4-1}   

\usepackage{graphicx}
\usepackage{setspace}
\usepackage{color}
\usepackage{float}
\usepackage{amsmath}
\usepackage{amssymb}
\usepackage{amsfonts}
\usepackage{dcolumn}
\usepackage{epsfig}
\usepackage{subfigure}
\usepackage{bm}
\usepackage{setspace}   
\usepackage{lipsum} 

\begin{document}

\title{Defect-Induced Photoluminescence of Strontium Titanate and its Modulation by Electrostatic Gating}
\author{Dushyant Kumar and R. C. Budhani*}
\affiliation{Condensed Matter Low Dimensional Systems Laboratory, Department of Physics, Indian Institute of Technology, Kanpur 208016, India}
\email{rcb@iitk.ac.in}
\date{\today}

\begin{abstract}

The photoluminescence (PL) spectra of Ar$^+$-ion irradiated single crystals of SrTiO$_3$ (STO) excited by 325 nm line of a He-Cd laser are compared with those of pristine crystals, epitaxial films and amorphous layers of STO at several temperatures down to 20 K. The 550 eV Ar$^+$-beam irradiation activates distinctly visible three PL peaks; blue ($\sim$430 nm), green ($\sim$550 nm), and infra-red ($\sim$820 nm) at room temperature making the photoluminescence multi-colored. The abrupt changes in PL properties below $\approx$100 K are discussed in relation with the antiferrodistortive structural phase transition in SrTiO$_3$ from cubic-to-tetragonal symmetry which makes it a direct bandgap semiconductor. The photoluminescence spectra are also tuned by electrostatic gate field in a field-effect transistor geometry. At 20 K, we observed a maximum increase of $\sim$20\% in PL intensity under back gating of SrTiO$_3$.

\end{abstract}
\maketitle

\section{Introduction}

Strontium titanate (SrTiO$_3$) is perhaps the most widely studied perovskite because of its unusual and technologically important properties, which also make it a promising material for the oxide-based electronics.\cite{cen2009oxide,kawasaki1994atomic,koster2000surface} At ambient temperature SrTiO$_3$ (STO) is a cubic crystal (Pm3m) with an indirect band gap of 3.27 eV. A cubic-to-tetragonal structural phase transition at $\approx$105 K makes STO a direct bandgap semiconductor and a precipitous growth of dielectric function follows below this temperature. High quality single crystals of STO have been used as substrates for epitaxial growth of many other transition metal oxides. It is also an ideal material for gate dielectric in field effect transistors (FET) due to its large dielectric constant.\cite{newns1998mott,nakamura2006low,rastogi2010electrically} The properties of STO can be varied from insulating to semiconducting\cite{Tufte1967semiSTO}, metallic\cite{Frederikse1964metSTO} and even superconducting at low temperatures\cite{schooley1964sup,koonce1967superconducting} on electron doping. Recently, it has been demonstrated that the interfaces of STO with perovskite oxides like LaAlO$_3$ and LaTiO$_3$ can stabilize a two dimensional electron gas (2DEG) with mobilities as high as $\sim$10$^4$ $cm^2/Vs$.\cite{ohtomo2004high,rastogi2012photoconducting,rastogi2010electrically} However, a 2DEG can also be formed on the surface of bare STO either by electrostatic gating or by Ar$^+$-ion irradiation.\cite{meevasana2011creation} The later creates oxygen vacancies on the surface of STO, which make it metallic.\cite{reagor2005ionmillSTO,kan2005blue} Many groups have studied this surface electron gas and have reported its unique properties\cite{Gentils2010PRB}, which include a large ($\approx$300\%) low temperature magnetoresistance\cite{bruno2011anisotropic}, electrostatic control of carrier concentration, persistent photoconductivity and its control by electrostatic gating.\cite{Dushyant2015PRB}

In the context of optical properties, while the undoped stoichiometric single crystals of STO do not show any PL at room temperature, a broad greenish luminescence is seen near $\approx$10 K. The intensity of this emission decreases rapidly above 60 K and disappears all together beyond 110 K.\cite{Grabner1969,sihvonen1967photoluminescence,Feng1982PRB,Leonelli1986PLSTO} The luminescence becomes pronounced when oxygen vacancies are incorporated in STO.\cite{kan2005blue,Sung2010PL100keV,Rho2009PLofannealedSTO,Rho2010PLofProtonSTO} Kan \textit{et al.} have noted that bombardment with 300 eV Ar$^+$ - ions induces blue ($\sim$420 nm) PL in stoichiometric single crystals of STO at room temperature, which they attribute to emission from oxygen vacancies related defect states.\cite{kan2005blue} However, a recent experiment of Sung \textit{et al.} on 100 keV Ar$^+$-ion irradiated STO shows only a broad luminescence centered at $\sim$510 nm at room temperature.\cite{Sung2010PL100keV} Since the nature of defects may change with ion energy, ion type and their fluence, these results indicate much diverse nature of defect physics in this material. The PL emission is generally derived from localized electronic states within the forbidden gap created by atomic vacancies or impurities.\cite{Szot1999surfaceRedOxiSTO,Cuong2007LDA+U} The vacancies/defects present in Ar$^+$-ion irradiated STO may not be of just single kind but in various different forms leading to far more abundant defect states. Indeed, the local density approximation (LDA) + Hubbard U study carried out recently on oxygen deficient STO predicts that linear vacancy clusters result in many localized in-gap states.\cite{Cuong2007LDA+U} This suggests the possibility to activate several PL emissions simultaneously at room temperature in Ar$^+$-ion irradiated STO. Moreover, it is expected that the nature and abundance of the surface states can be tuned by electrostatic gating to achieve multi-color optical devices. In view of these interesting predictions of the LDA+U theory, it is desirable to further explore the photoluminescence of Ar$^+$-ion irradiated STO and its other forms over a broad range of temperatures to cover the direct to indirect optical gap regimes. Moreover, since the PL is a very sensitive and selective probe of defect/impurity states, studies of luminescence would improve our understanding of the persistent photoconductivity reported earlier in Ar$^+$-ion irradiated STO.\cite{Dushyant2015PRB}

Here we present a detailed study of photoluminescence of oxygen-deficient SrTiO$_3$ created by Ar$^+$-ion irradiation over a broad range of temperature and compare it with our measurements of the PL in pristine crystals, epitaxial films and amorphous layers of the same material. While we did not observe any luminescence from the pristine STO at ambient temperature, the ion irradiation led to a multi-frequency emission. However, below $\simeq$105 K the characteristic PL of the unperturbed STO below the irradiated layer emerges and grows into a broad peak centered at $\approx$500 nm. It is also shown that the photoluminescence spectrum can be modulated by electrostatic gating at low temperatures ($<$20 K) which may be potentially important for applications.

\section{Experimental details}

The stoichiometric single crystals of SrTiO$_3$ used in this study were acquired from Crystal GmbH Germany. The (001) surface of these 0.5 mm thick and optically polished plates was irradiated at room temperature by Ar$^+$ - ions with a cumulative doses of $\sim$3.0$\times \:\:10^{18}$, 4.2 $\times \:\:10^{18}$ and 6 $\times \:\:10^{18}$ $ions/cm^2$. The typical acceleration voltage and ion current used in these experiments, carried out with a Kauffman type ion source operated at $\sim$8.5 $\times\,10^{-4}$ mbar Ar pressure, were 550 V and 1.5 $mA/cm^2$, respectively.
The irradiated surface of STO is metallic down to the lowest temperature ($\approx$10 K) where the sheet resistance and carrier mobility are $\sim$3 $\Omega$/$\Box$ and $\sim$0.2 $\times$ 10$^4$ cm$^2$/Vs respectively. Further details of the Ar$^+$-ion irradiation experiments and measurements of electronic transport in irradiated samples are given in our earlier article.\cite{Dushyant2015PRB} The photoluminescence spectra were excited with the
325 nm line of a He-Cd laser, and measured using Jobin Yvon Triax-320 spectrometer. For measurements of the temperature dependence of PL spectra, the samples were mounted in a close cycle helium cryostat having a quartz window for optical access. The PL response was also modulated electrostatically by gating the irradiated surface in a back gate configuration.

\section{Results}

\begin{figure}[h]
    \centering
    \subfigure{\includegraphics[ trim=0cm 0cm 0cm 0cm, width=7.5cm, angle=0 ]{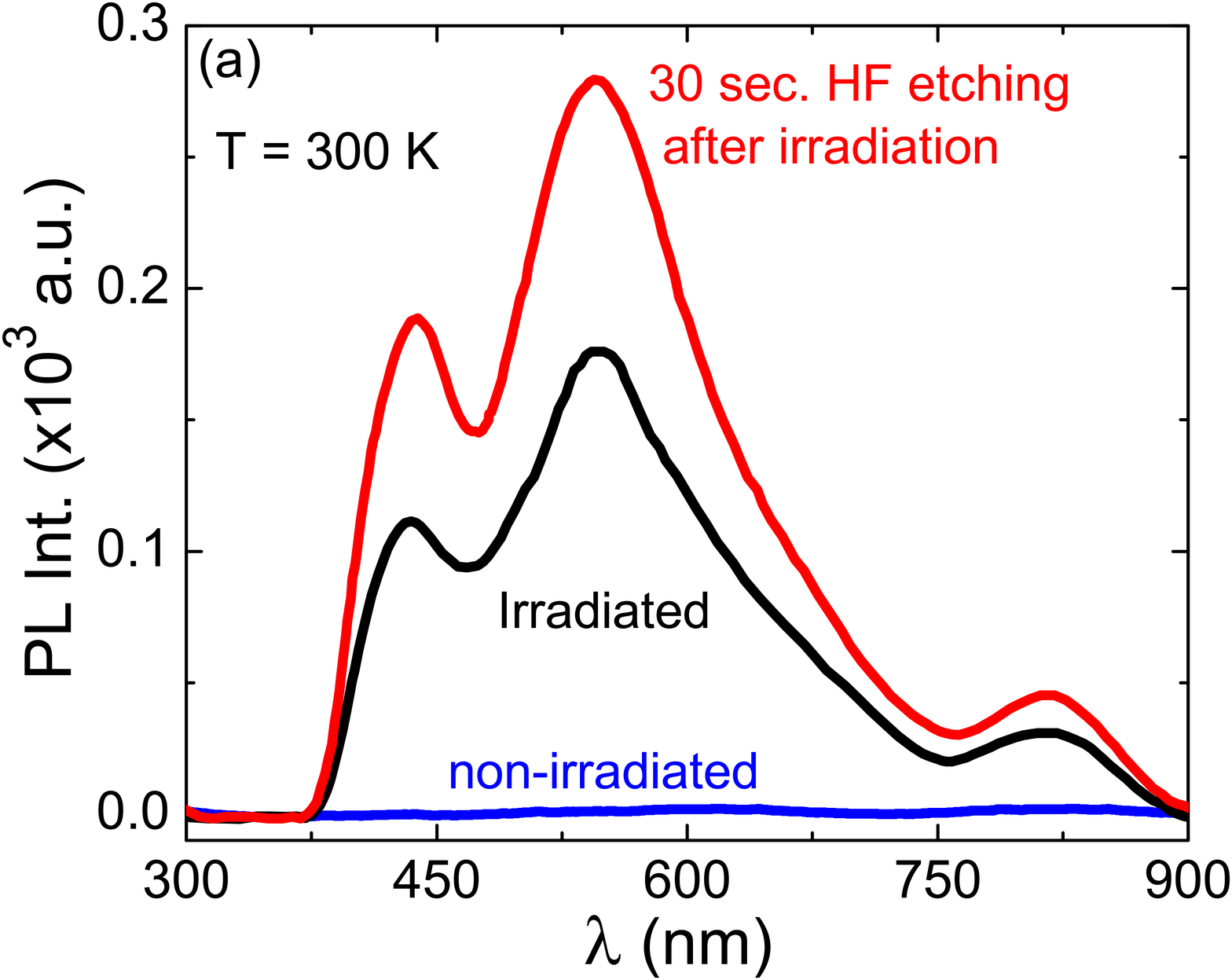}}
    \subfigure{\includegraphics[ trim=0cm 0cm 0cm 0cm, width=7.5cm, angle=0 ]{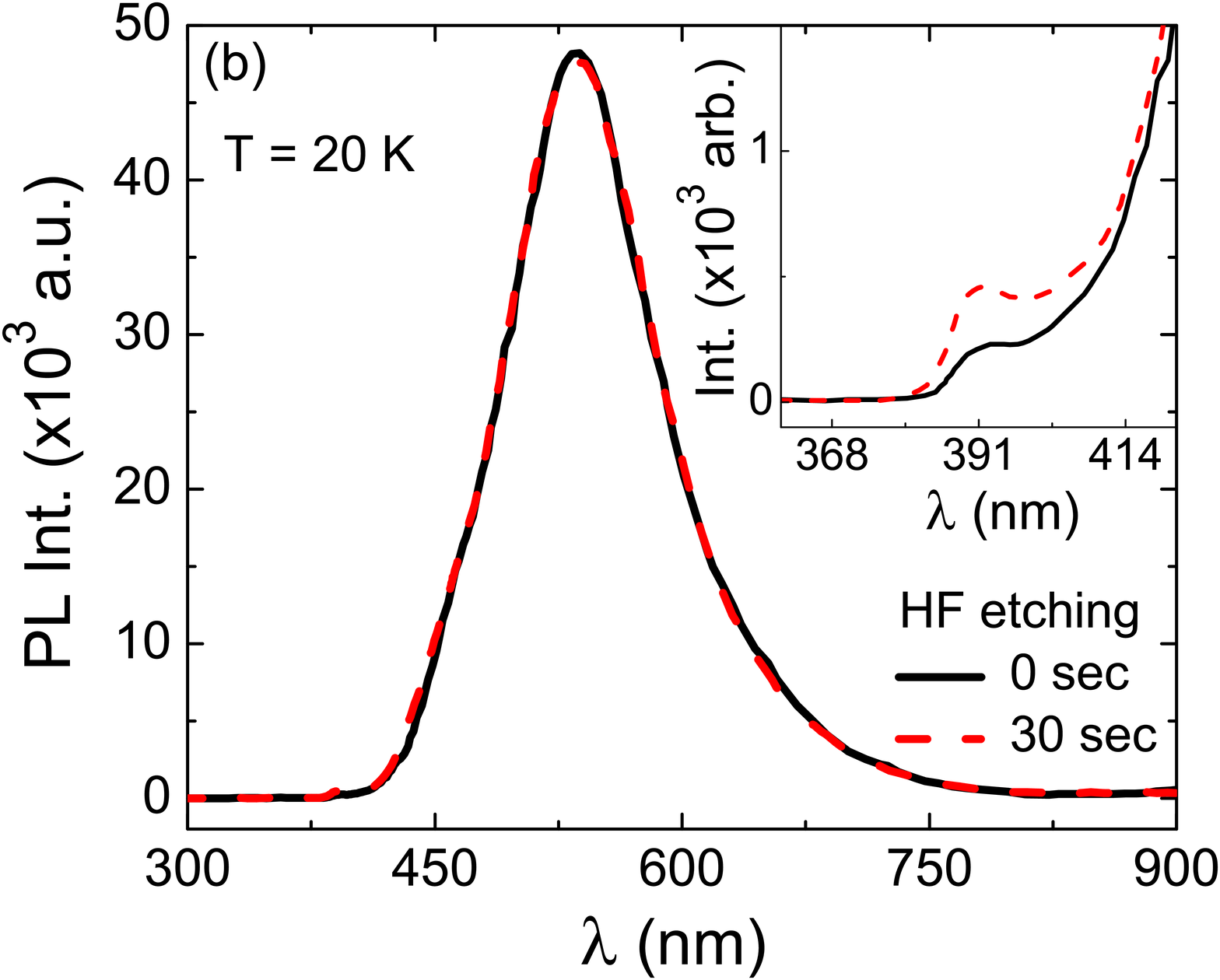}}
\caption{\label{fig 1} (Color online) Multi-color PL spectra of Ar$^+$-ion irradiated STO and the effect of hydrofluoric (HF) acid etching (a) at 300 K and (b) at 20 K. The HF treatment of irradiated STO increases the PL intensity at 300 K by a factor of two. For reference, the PL of non-irradiated stoichiometric STO is also shown in Fig. 1(a). The 390 nm peak at 20 K is zoomed in the inset of panel (b).}
\end{figure}

In Fig. 1(a), we show the room temperature photoluminescence spectrum of the (001) surface of Ar$^+$-ion irradiated SrTiO$_3$. In the same figure we have also shown the feature less PL response of a non-irradiated sample. The richness of the spectrum in the former case is a testimony of enhanced optical activity as a result of ion irradiation. The spectrum shows three distinct emission peaks centered at $\sim$430 nm, $\sim$550 nm and $\sim$820 nm, which on the energy scale correspond to $\sim$2.9 eV, $\sim$2.3 eV, and $\sim$1.5 eV, respectively. The blue luminescence ($\sim$430 nm) observed in this study is essentially the same as that reported by Kan \textit{et al.} in their 300 eV Ar$^+$-ion irradiated STO.\cite{kan2005blue} But these authors did not observe the two additional peaks seen here which result in a multicolored photoluminescence at room temperature in these 550 eV Ar$^+$-ion irradiated STO crystals.

Since heavy ion irradiation can cause amorphization of the target surface, we have also studied the effects of post irradiation etching of STO on its photoluminescence characteristics. For this purpose two STO samples 3 $\times$ 5 mm$^2$ were irradiated together to avoid any discrepancy. After irradiation, one of the samples was chemically etched in HF solution (NH$_4$F : DI water : HF = 18.54 gm : 50 ml : 3.75 ml) for 30 seconds. For comparison, the room temperature PL spectra of the etched sample is also plotted in Fig. 1(a). It is observed that the HF etching increases the integrated PL peak intensity by a factor of two keeping the position and shape of the peaks intact. The ion bombardment would create a large number of nanograins on the surface of the STO.\cite{li2007modulation} The SrO being very sensitive to HF attack, the etching process is likely to increase the amount of oxygen-vacancy defects by dissolving the SrO on the lateral sides of these nanograins\cite{li2007modulation} thus raising the conduction band carrier density and hence the PL intensity.

\begin{figure}[h]
\begin{center}
\includegraphics [ trim=0cm 0cm 0cm 0cm, width=8.5cm, angle=0 ]{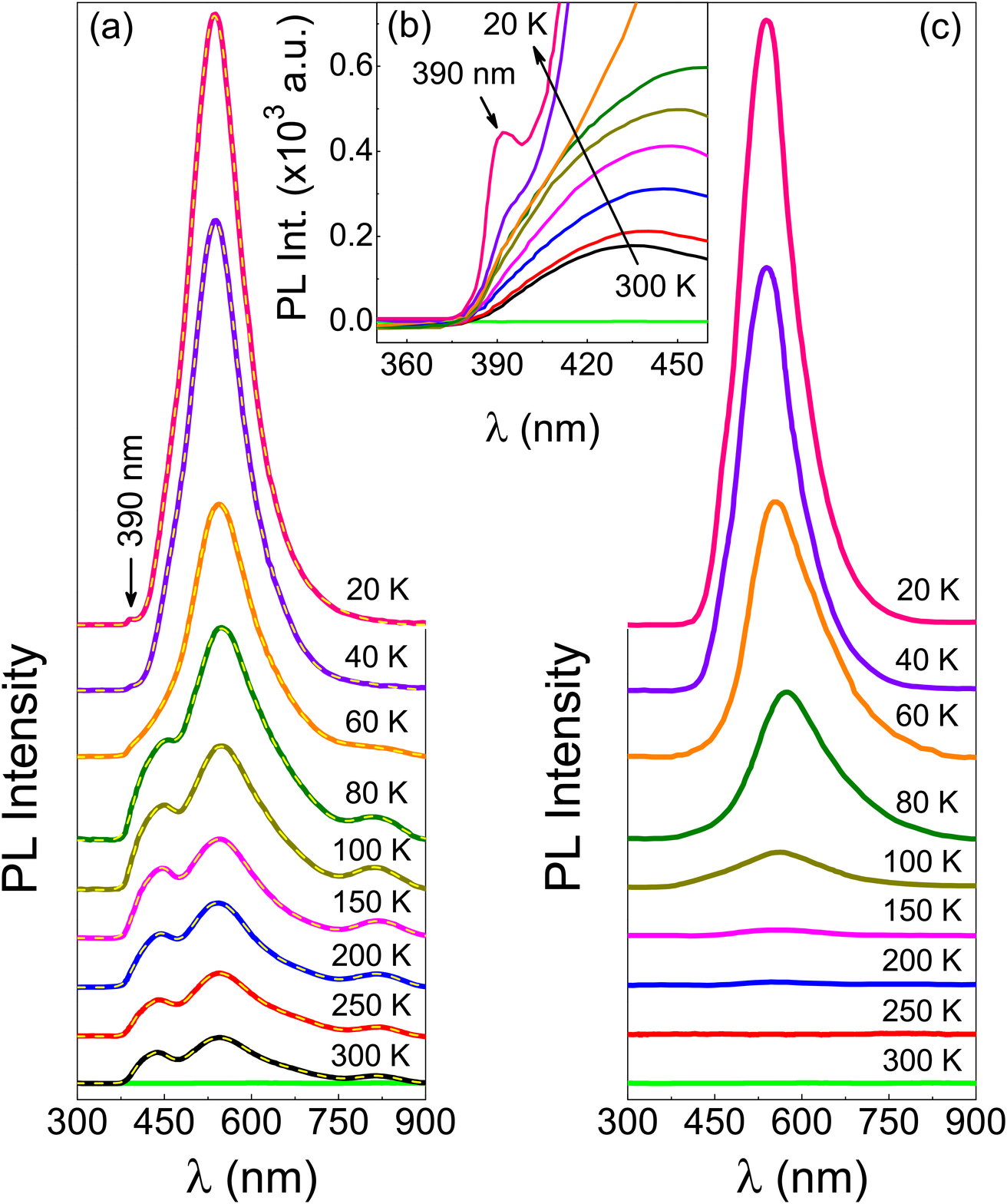}%
\end{center}
\caption{\label{fig 2} (Color online) PL spectra of Ar$^+$-ion irradiated STO and non-irradiated (bare) STO in the temperature range of 300 K to 20 K. Panel (a) shows the temperature dependence emission spectra of Ar$^+$-ion irradiated STO. For clarity, the spectra along with their deconvolution fit are shifted and the intensity of PL spectra at 20 K, 40 K and 60 K are reduced by factor of 13, 10 and 3. These spectra are put together in panel (b) and zoomed in the vicinity of 390 nm peak emphasizing its appearance below $\sim$60 K. The similar measurements carried out for bare STO are shown in panel (c) for comparison. The spectra at 20 K and 40 K are reduced by factor of 12 and 9.}
\end{figure}

Figure 2 illustrates the changes in PL spectra of the irradiated and etched samples at various temperatures from 300 K down to 20 K. From panel (a), it can be seen that the intensity of all the PL peaks increases gradually on decreasing the temperature down to $\sim$100 K. A slight shift in the position of the blue peak towards higher wavelength can also be noticed on cooling. On further decreasing the temperature from 100 K down to 60 K, the green luminescence intensity increases dramatically, whereas, the blue luminescence appears to merge under the now much broader emission with its peak at green. A closer examination of these spectra further reveals the presence of another emission peaked at $\sim$390 nm [marked by an arrow in Fig 2(a) and Fig. 2(b)] whose intensity increases on lowering the temperature. On reaching 20 K, an intense broad luminescence ranging from $\sim$370 nm to $\sim$850 nm along with this extra peak centered at $\sim$390 nm is clearly visible in the figure [panel (a) and (b)].
Note that the later is not observed in stoichiometric non-irradiated STO [panel (c)]. The PL spectrum of un-etched (only irradiated) STO has also been collected at 20 K. The spectrum is shown in Fig. 1(b) along with that of the etched irradiated STO. One can see that there is no effect of etching on the greenish broad band luminescence, however, as clear from the inset of Fig. 1(b), it increases the 390 nm PL peak intensity by a factor of two.

\begin{figure}[h]
\begin{center}
\includegraphics [ trim=0cm 0cm 0cm 0cm, width=7.5cm, angle=0 ]{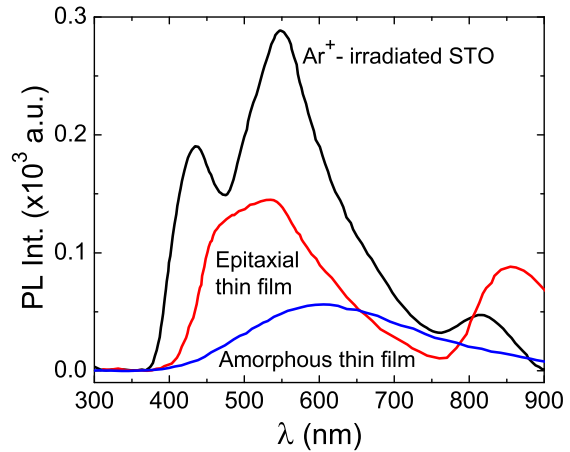}%
\end{center}
\caption{\label{fig 3} (Color online) Photoluminescence profile of epitaxial oxygen deficient thin film and amorphous film of SrTiO$_3$ at room temperature. For comparison, the PL spectra of Ar$^+$-irradiated STO is also shown. Note that the amorphous film shows a broad featureless emission profile whereas the PL spectra of oxygen deficient film behave similar to that of Ar$^+$-irradiated STO.}
\end{figure}

 \begin{figure}[h]
\begin{center}
\includegraphics [ trim=0cm 0cm 0cm 0cm, width=7.5cm, angle=0 ]{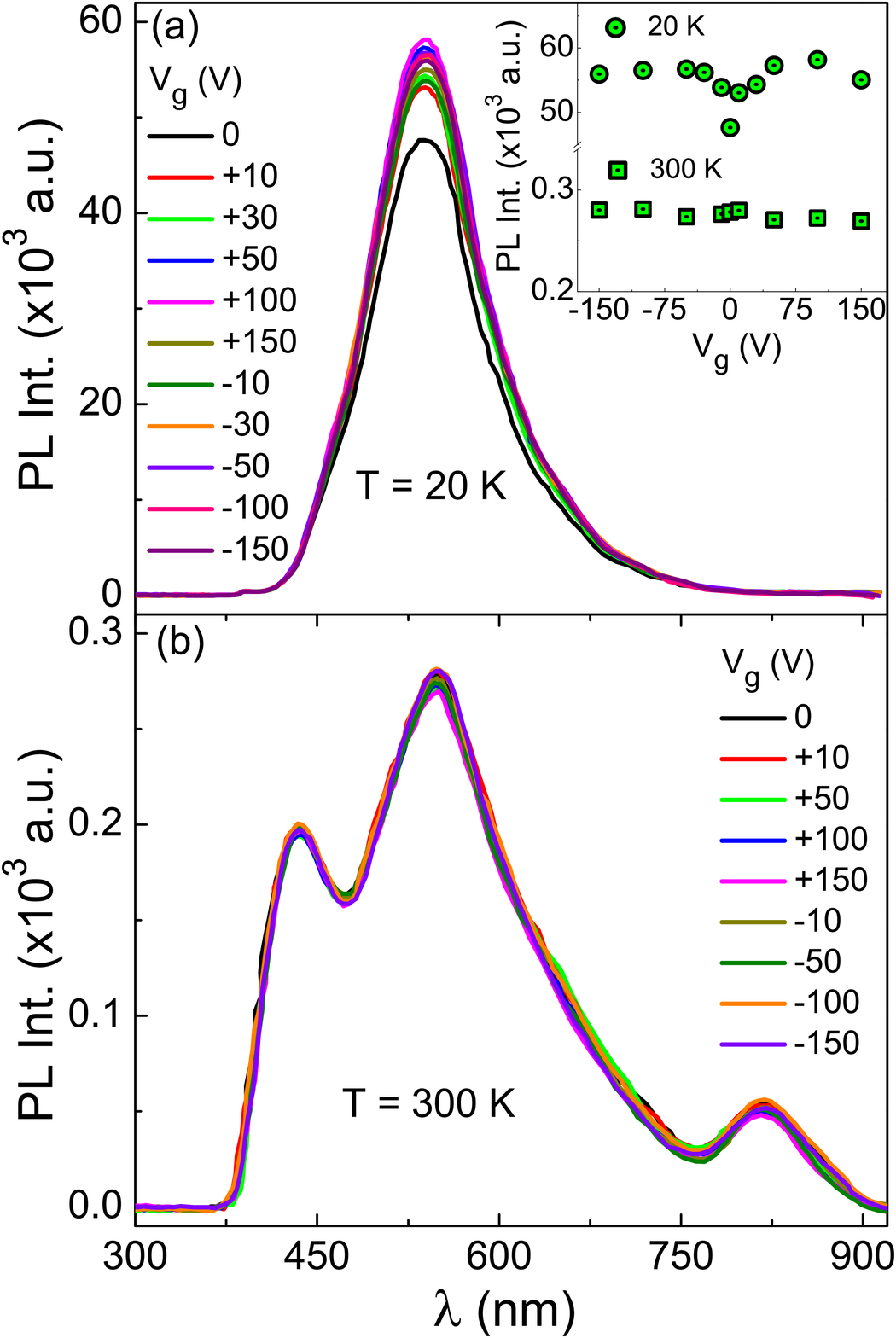}%
\end{center}
\caption{\label{fig 4} (Color online) PL modulation by electrostatic gate field varying from -150 V to +150 V (a) at 20 K and (b) at 300 K. The changes in broad green peak intensity as a function of gate field are shown in the inset of panel (a).  At 20 K, the gate field increases the peak-intensity where as there are no noticeable changes at 300 K.}
\end{figure}


We now turn our attention to the possible sources of the rich spectrum, whether it is oxygen deficient layer or the top amorphous layer. Towards this end, we have examined the PL properties of oxygen deficient epitaxial thin films as well as amorphous films of STO. These 100 nm thick films were grown on STO (001) substrates by pulsed laser ablation of a bulk target of SrTiO$_3$. The oxygen deficient film was deposited under reduced oxygen environment (6.3 $\times$ 10$^{-6}$ mbar) at 800$^0$C substrate temperature. The epitaxial growth was confirmed by X-ray diffraction. This film showed a metallic behaviour down to 10 K coming from oxygen vacancy induced mobile electrons in the system as reported earlier by Perez-Casero \textit{et al.}\cite{perez2007thin} To grow the amorphous film, the deposition was carried out at room temperature under 1 $\times$ 10$^{-4}$ mbar of oxygen. Two samples of each type were prepared. Both of them showed similar PL results. The room temperature PL spectra of these films are shown in Fig. 3 along with those of the Ar$^+$-irradiated STO crystal. The amorphous film displays a broad featureless luminescence ranging from 350 nm to 950 nm, which is in accordance with the reported photoluminescence of amorphous SrTiO$_3$.\cite{soledade2002amorSTO,pinheiro2003defectSTO,orhan2004defectSTO} This emission profile clearly does not match with the luminescence of Ar$^+$-irradiated STO. However, the emission behaviour of the oxygen deficient epitaxial film is close to that of the Ar$^+$-irradiated STO, which suggests that the photoluminescence in the later is mainly originating from the oxygen deficient layer.

A unique feature of the present study is the modulation of the PL spectra with an electrostatic gate field as shown in Fig. 4(a) under a gate voltage varying from -150 V to +150 V at 20 K. These voltages translate into an electric field of -3 kV/cm to +3 kV/cm. The entire sweep of gate field from zero to positive to negative and then back to zero took around $\sim$2 hours. Over this time scale, the data are reversible. As seen in the figure the broad green peak intensity increases on gating symmetrically without any noticeable change in the peak position. The inset of Fig. 4(a) shows the peak intensity as a function of gate voltage within the range of -150 V to +150 V. The response is unipolar. We did not observe any gate field dependence of the PL spectra at 300 K [see Fig. 4(b)].

\section{Discussion}

The broad spectrum stretching from $\sim$380 nm to $\sim$ 850 nm with three distinct peaks suggests that there may be additional specific features which are hidden under the envelope. To better understand the PL spectrum and its origin, we have analyzed the emission collected at 300 K by deconvolution using gaussian line-shapes. Multiple peaks which account for the measured intensity profile are shown in Fig. 5. Table-I lists the position of these peaks and their spectral weight.

\begin{figure}[h]
\begin{center}
\includegraphics [ trim=0cm 0cm 0cm 0cm, width=7.5cm, angle=0 ]{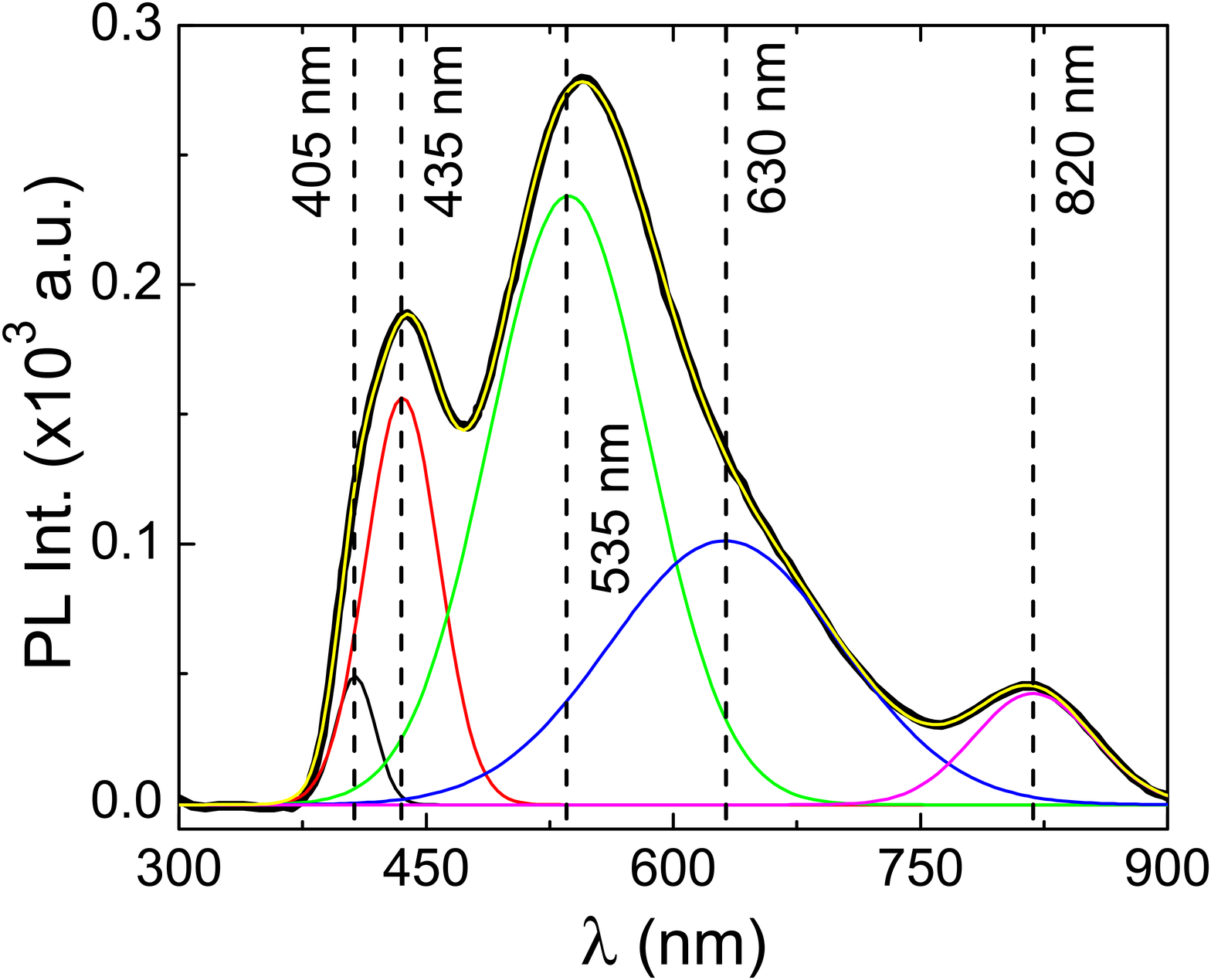}%
\end{center}
\caption{\label{fig 5} (Color online) Deconvolution of PL spectra observed at room temperature. It reveals the presence of two more peaks along with three peaks, which are distinctly visible in the spectra. Multiple frequencies are marked by dashed lines. Cumulative fit is also drawn.}
\end{figure}

\begin{figure}[h]
\begin{center}
\includegraphics [ trim=0cm 0cm 0cm 0cm, width=8.25cm, angle=0 ]{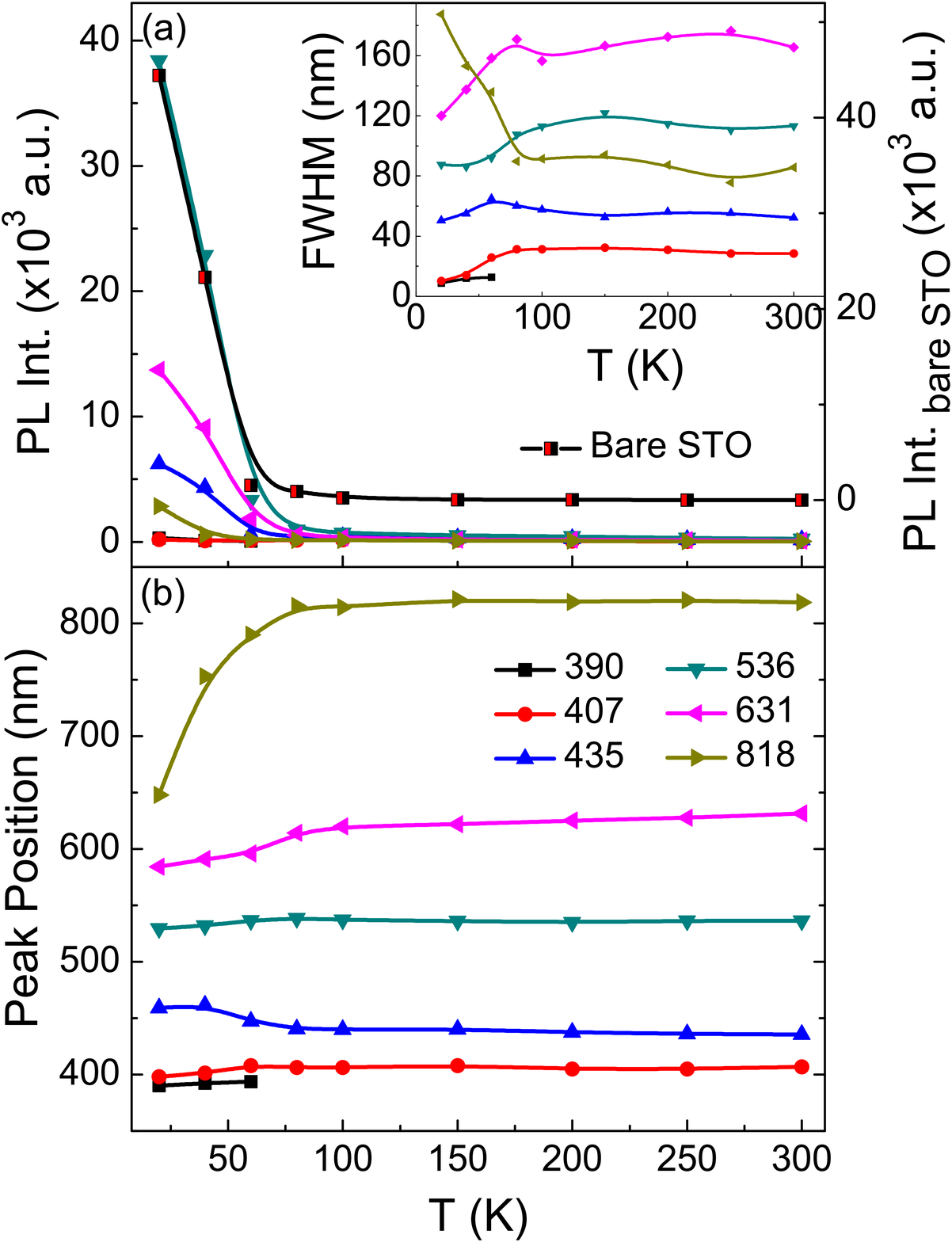}%
\end{center}
\caption{\label{fig 6} (Color online) Temperature dependence of the characteristic features of the multiple peaks; (a) intensity, (b) peak position, and the inset of (a) shows the FWHM. These features were extracted from the deconvolution of the PL spectra collected at several temperatures between 300 K to 20 K. Drastic changes in all of three can be seen around $\lesssim$100 K. For comparison, the intensity of greenish luminescence of bare STO is also plotted in panel (a). Lines are guide to the eye.}
\end{figure}


\begin{table}[h]
\begin{center}\caption{The percentage area of each decomposed PL peak at 300 K.} \label{TABLE I.}
\vskip 1cm
\begin{tabular} {|c|c|c|c|c|}
\hline
Violet & Blue & Green &
Red & Infra-red \\
\hline
407 nm & 435 nm & 536 nm &
631 nm & 818 nm \\
\hline
3\% & 15\% & 47\% &
30\% & 7\% \\
\hline
\end{tabular}
\end{center}
\end{table}

A similar analysis has been done for all the spectra collected in the temperature range of 20 K to 300 K. The cumulative fits are drawn in Fig. 2(a). The peak positions, their full width at half maximum (FWHM) and intensity derived from the deconvolution procedure are plotted in Fig. 6 as a function of temperature. The intensity of all the peaks increases gradually on lowering the temperature from 300 K to 100 K. This trend accelerates on further decreasing the temperature below $\approx$100 K. Two different mechanisms appear to be responsible for the luminescence as indicated by the two distinct regions of behaviour well separated at $\approx$100 K. From 300 K to $\approx$100 K, the blue and red peaks shift gently towards green with no noticeable shift of the IR and violet peak positions.
However, in low-T region (below $\approx$100 K), the response is remarkably different. At 60 K, the diverse nature of luminescence is no more visible. There is only a broad emission profile peaked at the position of the green line. It seems that all other peaks are merged under this broad and intense luminescence. Moreover, the intensity of this green signal increases sharply on lowering the temperature. In the low T-region, a similar T-dependence has been reported for the greenish luminescence observed in undoped-stoichiometric STO.\cite{Grabner1969,Feng1982PRB} We have also recorded the PL spectra of stoichiometric non-irradiated STO (001) substrate throughout the temperature range of 300 K to 20 K. The results are shown in panel (c) of Fig. 2. At 20 K, one can see an intense broad emission peaked at 540 nm, whose intensity decreases rapidly as the temperature is raised upto 100 K, and becomes negligibly small at 150 K with no detectable signal beyond 200 K. The PL spectra of non-irradiated STO at 20 K resembles well with that of the irradiated sample [Fig. 2(a)] except that the latter shows an extra peak in the vicinity of 390 nm. Moreover, the T-dependence of the green peak intensity of the irradiated sample in low T-region agrees well with that of the non-irradiated STO [panel (a) of Fig. 6]. This comparative study suggests that below $\approx$100 K the luminescence of the non-irradiated part of STO underneath the disordered region created by Ar$^+$-ion irradiation dominates over the emission signal coming from the oxygen deficient layer in the ion affected region. Note that all of the PL properties (peak position, line width and PL intensity) change abruptly below around $\approx$100 K, the temperature where SrTiO$_3$ crystal undergoes a structural phase transition from cubic-to-tetragonal.

The multicolored PL seen in these experiments is consistent with the predictions of LDA+U calculations of Cuong \textit{et al.}\cite{Cuong2007LDA+U} on oxygen deficient STO, where they have shown that a single oxygen vacancy can create a shallow level just below ($\sim$0.11 eV) the conduction band minimum of STO. This will lead to emission at $\approx$390 nm. Incorporation of vacancy-vacancy interactions in the calculation yields oxygen vacancy clusters which induce localized electronic levels ranging in energy from 0.3 eV to 1.14 eV in the forbidden gap with respect to the conduction band edge. Defect clusters can also result from binding of oxygen and strontium vacancies. The accelerated Ar$^+$-ions, while penetrating into the STO lose their energy resulting in an oxygen-deficient layer along with an amorphous layer on the top.\cite{kan2005blue} We believe that the ion irradiation is creating disorder in the material leading to several different kinds of oxygen-vacancy clusters which induce many in-gap localized states. The recombination of electrons trapped in these levels to the valence band holes can lead to multi-frequency emission at room temperature.

At this point it is worth commenting on why the pristine STO crystal shows an onset of photoluminescence at $\sim$100 K which becomes quit pronounced around $\simeq$20 K. It is well known that cubic (Pm3m) SrTiO$_3$ undergoes an antiferrodistortive ($O_h^1$ $\rightarrow$ $D_{4h}^{18}$) structural phase transition (AFD-PT) to a tetragonal symmetry (I4/mcm) in the vicinity of 105 K.\cite{shirane1969lattice} In the cubic STO, the top of the valence band is located at the R-point in the first Brillouin zone where as the lowest conduction band bottom lies at the $\Gamma$-point making this transition indirect. The AFD-PT results in merging of the R and $\Gamma$-points of the Brillouin zone of the cubic structure into the $\Gamma$-point of tetragonal phase, thereby transforming the lowest in energy, indirect phonon-assisted R $\rightarrow$ $\Gamma$ optical transition into the direct $\Gamma$ $\rightarrow$ $\Gamma$ transition.\cite{heifets2006calculations,fleury1968soft,mattheiss1972effect} We believe that this opening of the direct gap results in enhanced PL intensity in the pristine STO below 100 K.

Now we discuss the possible origin of the increase in PL intensity as seen on applying a gate field. A voltage (positive/negative) on the gate induces onto the sample a charge (negative/positive), which, in the absence of surface states, distributes throughout a space charge region $\lambda$ either in gap states or in the bands. Using Barbe theory of field effect\cite{barbe1971theory}, the induced space charge (Q$_{sc}$) per unit area in the semiconductor can be expressed as
\begin{equation*}
\mid Q_{sc}\mid\,\, = \Big\{\frac{\epsilon \epsilon_0 kT} {e\lambda}\Big\}y_s
\end{equation*}
where $\epsilon$ is the dielectric constant of the material, $\epsilon_0$ is the permittivity of free space, k is the Boltzmann constant and $y_s$ is the dimensionless energy at the surface.
However, in the presence of surface states/trapping centers in the material, most of the induced charge falls into these states. The charge (Q$_{ss}$) residing in surface states can be written as
\begin{equation*}
\mid Q_{ss}\mid\, = ekT(N_{AS}^F + N_{DS}^F)y_s
\end{equation*}
where $N_{AS}^F$ and $N_{DS}^F$ are the number of acceptor type and donor type surface states per unit area per unit energy at the Fermi level, respectively.

In our case, the ion irradiation process is likely to create a large number of defects and thus midgap energy states. These states are plausible to be hole as well as electron trap levels forming non-radiative recombination sites which lower the intrinsic PL of STO in the direct band gap state.\cite{kang2009high} These positive and negative recombination centers can efficiently be deactivated by the variation of relative position of the Fermi level using negative and positive gate fields respectively\cite{galland2011two,kang2009high,li2014enhanced,schornbaum2015light}. We expect the filling of electron trap states while the gate field is swung positive and the passivation of hole trap states when the gate field is swung negative. Therefore, the gate field regardless of its polarity will ensure the reduction of non-radiative recombination and hence the increase in PL intensity. On the other hand, at 300 K, where the gate field due to low dielectric function is not that effective, it can only partially passivate these centers and hence no significant modulation of PL intensity is seen.


\section{Conclusions}

In summary, we have shown that the 550 eV Ar$^+$-ions irradiation induces a multi-color photoluminescence in stoichiometric single crystals of SrTiO$_3$ (001) at room temperature, where no PL signal is seen in the pristine state. The PL spectra of oxygen deficient and amorphous thin films of SrTiO$_3$ suggest that this multi-color emission originates from oxygen deficient layers of the irradiated crystal. The luminescence intensity is enhanced further when the ion induced amorphous layer on surface of the STO is removed by etching. On cooling from 300 K to 100 K, the luminescence intensity increases with a noticeable shift of the blue and red peaks toward the middle of the spectrum. The diverse nature of PL disappeared below 80 K with a drastic increase in green luminescence. At 20 K, a modulation of PL spectra was achieved through the passivation of surface states by electrostatic back gating of the STO.

\begin{acknowledgements}
The authors thank A. Rastogi for initial help while doing photoluminescence measurements. D.K. would like to acknowledge Indian Institute of Technology Kanpur for partial financial support. R.C.B. acknowledges J. C. Bose National Fellowship of the Department of Science and Technology, Government of India.
\end{acknowledgements}

\bibliography{references}

\end{document}